\documentclass[%
 reprint,
 amsmath,amssymb,
 aps,nofootinbib,   
]{revtex4-1}
\usepackage{bm}
\PassOptionsToPackage{linktocpage}{hyperref}
\usepackage[hyperindex,breaklinks]{hyperref}
\usepackage{enumitem}
\usepackage{slashed}

\renewcommand{\theta}{\vartheta}

\newcommand{\ket}[1]{\ensuremath{\left|\, #1\right>}}

\usepackage{array}
\usepackage{mathtools}

\usepackage{etoolbox}

\begin{document} 

\title{$S$-Matrix and Anomaly of de Sitter}

\author{Gia Dvali  
} 
\affiliation{%
Arnold Sommerfeld Center, Ludwig-Maximilians-University,  Munich, Germany, 
}%
 \affiliation{%
Max-Planck-Institute for Physics, Munich, Germany
}%


\date{\today}

\begin{abstract}

$S$-matrix formulation of gravity 
 excludes 
 de Sitter vacua. In particular, this is organic to string theory. 
  The $S$-matrix constraint is enforced by an anomalous 
  quantum break-time proportional to the inverse values of 
  gravitational and/or string couplings. 
 Due to this, de Sitter can satisfy the conditions for a valid vacuum only at the expense of
 trivializing the graviton and closed-string $S$-matrixes. 
  At non-zero gravitational and string couplings, 
 de Sitter is deformed  by corpuscular $1/N$ effects, similarly to 
 Witten-Veneziano mechanism in QCD with $N$ colors. 
In this picture, an $S$-matrix formulation of Einstein gravity, such 
as string theory, nullifies an outstanding cosmological puzzle.
 We discuss possible observational signatures which are especially interesting 
 in theories with large number of particle species. 
 Species can enhance the primordial quantum imprints to potentially  observable level even if the standard inflaton fluctuations are negligible. 

 \end{abstract}

\maketitle

\subsection{Introduction} 
 
  String theory is the most prominent example of a theory based on $S$-matrix. 
 This formulation demands the existence of a valid $S$-matrix vacuum
 such as Minkowski. It also excludes the de Sitter 
 space-time from the list of the valid vacuum states.
 The first indication of this appears already at the classical level. 
 It comes from the absence of a globally-defined time 
 in de Sitter space. \\
 
  Can a de Sitter vacuum exist in quantum theory? 
   Following \cite{Dvali:2013eja, Dvali:2014gua, Dvali:2017eba},
  we shall give arguments indicating that in a theory with
  interacting gravitons (closed strings) this is not possible.   
  \\

 In classical General Relativity (GR), the de Sitter metric is 
 sourced by a positive vacuum energy density, $\Lambda$, which 
 we call cosmological constant (or a cosmological term). This parameter 
 is highly sensitive to a cutoff of the theory. 
 This is the essence of the celebrated cosmological constant puzzle. 
 It is usually viewed as the problem of fine-tuning or naturalness. 
 The solution is often attributed to an anthropic selection of our  
vacuum with small $\Lambda$ on a vast string landscape of 
plentitude of de Sitter vacua with various energy densities. 
 The arbitrariness accompanying this proposal 
is sometimes used as the point of criticism as lacking predictivity.
  \\
  
  The purpose of the present note is to  argue that the situation is exactly the opposite.  If there is any quantity that string theory severely constraints, it is 
 $\Lambda$.  
 According to \cite{Dvali:2013eja, Dvali:2014gua, Dvali:2017eba}, de Sitter vacua are inconsistent with quantum gravity.
  The problem with positive $\Lambda$ is not a matter 
 of fine tuning but of consistency. 
  The reason is 
 that such ``vacua" exhibit an anomalous {\it quantum break-time}, $t_Q$,   
 incompatible with the notion of a vacuum. 
  The phenomenon of {\it quantum breaking} amounts to  
 a full departure from the classical description. Such a departure 
 is evidently in conflict with a stationary (or slow-varying) classical source, such as $\Lambda$. This has number of profound implications.
In particular,  $\Lambda$ cannot be a source of the 
energy density in the Universe. 
  \\

  A formulation of quantum gravity in de Sitter space
  is connected with well-known difficulties 
  (see, \cite{Witten:2001kn}  
  and references therein).  
  The approach of \cite{Dvali:2011aa, Dvali:2013eja, Dvali:2014gua, Dvali:2017eba}, 
  changes the perspective on the problem. Instead of treating de Sitter 
  as {\it vacuum}, it is viewed as a coherent state of gravitons 
  constructed on top of a Minkowski vacuum. 
   This view gives one a double 
  advantage. \\
  
  First, whatever happens to (or in)  de Sitter, is now a part of 
  a quantum performance staged on Minkowski.  Of course, 
  nothing is free of technicalities, but  such a performance can, 
 at least in principle,  be traced.  Perhaps, the optimal 
characteristics 
can be found in terms of what Witten 
 calls meta-observables \cite{Witten:2001kn}. \\
 
 The second bonus is that a coherent state description of de Sitter 
 imposes upon us its corpuscular resolution in terms of $N$ constituent 
 gravitons, the picture introduced for 
 de Sitter, as well as for black holes, in \cite{Dvali:2011aa}.
 This resolution reveals phenomena that are 
 fundamentally impossible to detect when we treat de Sitter as 
 a vacuum. 
   In particular, the anomalous quantum break-time is caused by 
 $1/N$ effects \cite{Dvali:2013eja, Dvali:2014gua, Dvali:2017eba}. 
 The vacuum limit of de Sitter corresponds to $N \rightarrow \infty$, 
 for which the corpuscular effects vanish.  
   In a well-defined sense, these $1/N$ effects are analogous
 to the ones in QCD with 
 $N$ colors.  There too, $1/N$ corrections vanish in the 't Hooft's planar limit 
\cite{tHooft:1973alw}.
 \\

 The present note is an attempt to substantiate this 
 result from a different angle.  We wish to point out that the quantum break-time  $t_Q$ represents  
 a measure of $S$-matrix inconsistency of a de Sitter-like state. 
 The catch is that $t_Q$ scales as inverse of the gravitational (or string) 
 coupling. Correspondingly, the parameter  choice
 that supports $t_Q=\infty$, decouples gravity.  \\

  It is evident that the above link is universal for 
 any $S$-matrix theory of gravity. In particular, it is applicable 
 to string theory, at least at weak string coupling.    
  For any given finite value of the curvature radius, $R_{dS}$, 
  the gravitational quantum break time is proportional to $N$ and inversely 
  proportional to the gravitational coupling $G$.
 In string theory, $G$ can be expressed 
 through the string scale $M_s$ and the string coupling $g_s$. 
 So, in string theory, for a finite value of the string scale, $M_s$, the  anomalous quantum break-time scales as, 
   \begin{equation} \label{DSS}
     t_Q  \propto   N \propto \frac{1}{g_s^2} \,, 
    \end{equation}
  where the proportionality coefficients  are the functions of  $R_{dS}$,
 $M_s$ and the number of the light particle species. 
 These shall be made explicit later.  \\

 De Sitter can serve as a valid {\it vacuum}  only for $t_Q=\infty$.
From (\ref{DSS}), this requires $G=g_s=0$ ($N=\infty$), implying that 
the closed string and graviton $S$-matrixes are trivial.  
On the other hand, for  non-zero values 
of $g_s$ and $G$ (finite $N$), the system must exit gracefully from the de Sitter state 
before the time $t_Q$ elapses.   
 This imposes a general constraint that 
a cosmological 
  source must evolve faster than the quantum break-time.  \\
   
  In this way,  thanks to its $S$-matrix formulation,
  string theory nullifies an
outstanding cosmological puzzle. The same must remain  
true in any effective formulation of Einstein gravity based on $S$-matrix, 
regardless the nature of UV-completion. 
  \\

The inconsistency of de Sitter can be described in the language of a quantum
 anomaly.  One may say that the classical symmetry 
of de Sitter state is explicitly 
 broken by $1/N$  corpuscular effects.  However, in order 
 to avoid confusion, the meaning of this statement shall be explained 
 very carefully.  Usually, the  {\it anomaly} is associated with the breaking 
 of a symmetry of the Hamiltonian. This breaking may also deform 
 the vacuum.  In our $S$-matrix treatment, the de Sitter  is a state 
 on Minkowski \cite{Dvali:2013eja}.  It is mistaken 
 for the ``vacuum" by an observer that is blind 
 to $1/N$ effects. It is the symmetry of 
 this pseudo-vacuum that is broken by $1/N$ corrections.  \\
 
 In some aspects, this breaking of de Sitter symmetry 
 is spiritually similar to anomalous breaking of axial symmetry by  
 Witten-Veneziano  mechanism \cite{Witten:1979vv}
  in 
 QCD with $N$ colors.  In that case, $1/N$ effects generate a bias 
 that gives a finite decay-time to a pseudo-Goldstone boson of axial 
 symmetry. Similarly, in case of de Sitter,  $1/N$ effects lead to 
 an anomalous quantum break-time of de Sitter.  
 In both cases 
the critical time is determined by $N$.    
 After this time, de Sitter
is no longer able to satisfy a classical equation with the initial source. 
Thus, in order to maintain the validity of a classical approximation, 
necessary for the vacuum treatment,  the 
source must relax towards the asymptotic $S$-matrix vacuum. 
 \\

 The presented particle physicist's way of thinking is 
 somewhat alien for the way we are used to treat various backgrounds  
in classical GR. After all,  in GR,  Minkowski is one out of infinitely many possible 
choices of the metric.  Naively, it is not more special than, let us say,  
a de Sitter 
space.  Therefore,  from the perspective of GR, 
the de Sitter could be viewed as an equally good vacuum.  However, this equality is false.  Classically, the Minkowski vacuum represents 
a limit of de Sitter for $\Lambda \rightarrow 0$. However, from 
the $S$-matrix perspective the limit is not smooth.
For $\Lambda > 0$, no matter how small, a de Sitter state 
is subjected to a  finite anomalous quantum break-time.  
\\

\subsection{Double-Scaling Limit}

  As the first indication of discontinuity we offer the following scaling 
 argument.
  The argument works in arbitrary number of dimensions 
  but  we take $3+1$ first. Let us assume that a fundamental 
  theory of gravity gives us a de Sitter metric sourced by a constant 
  $\Lambda$.  The corresponding curvature radius is given by
  (irrelevant  numerical factors shall be ignored from now on),  
   \begin{equation} \label{Radius}
    R^{-2}_{dS} = \Lambda G \,, 
    \end{equation}  
where $G$ is the Newton's gravitational constant.  \\

  Let us imagine that we wish to describe some quantum gravitational  scattering process on such a vacuum. 
  For example, a $2\rightarrow 2$ graviton-graviton scattering. Gravitons of some characteristic wavelength $\lambda$ (or 
  effective momentum-transfer  $p \sim 1/\lambda$)
  interact via a quantum gravitational coupling,  
  \begin{equation}\label{AlphaGR}
  \alpha_{gr} \equiv p^2G = \frac{G}{\lambda^2} \, . 
   \end{equation} 
 Of course, since there is no global observer, the complete $S$-matrix 
  is impossible. However, we may attempt to replace it by some effective $S$-matrix description.    
  Indeed, if $\lambda \ll R_{dS}$, it looks like that the $S$-matrix formulation should be fine, at least approximately.  
   However, this requires the 
existence of a {\it vacuum} that is not affected by the scattered 
particles.  This may look like a small detail but it is crucial for a consistent 
$S$-matrix description of the process.  Putting it differently, 
the vacuum should not be able to ``recoil" and thus absorb some information. 
 In Minkowski, this is guaranteed by an infinite spread 
and the existence of a global time.  The prominent role 
of Minkowski shall be sharpened later. 
\\ 
 
   In order to achieve the same in de Sitter, we must pay a price. Namely,  
    we must 
  take the limit of a {\it rigid geometry} in which the {\it quantum} 
 back-reaction on de Sitter from the scattering particles vanishes. 
  Such a limit is unique: 
    \begin{equation} \label{GLimit}
    \Lambda \rightarrow \infty,~~G \rightarrow 0,~~
    \Lambda G = R^{-2}_{dS} = {\rm finite}.
    \end{equation}  
  However, in the very same double-scaling limit, the quantum coupling
  between the gravitons of any finite wavelength vanishes,  
   \begin{equation} \label{ALimit}
    \alpha_{gr} =  \frac{G}{\lambda^2}  \rightarrow 0\, .
 \end{equation}  
 Thus, we observe that in the unique limit in which de Sitter can be 
 consistently treated as vacuum, the scattering among gravitons vanishes. 
 In other words, in the {\it vacuum-limit}  of de Sitter, gravitons decouple: 
 \begin{equation} 
 ({\rm de \,  Sitter = vacuum}) ~ \rightarrow ~ \alpha_{gr} =0 \,.    
\end{equation}  
 
 Notice that in the same limit, there is no problem of keeping 
  any non-gravitational interaction, e.g., 
 the electroweak interaction, to have a 
 finite strength. So the issue is mainly connected with the quantum 
 gravitational part.   This simple scaling argument shows why quantum gravity is special and why it is directly linked with the value of  $\Lambda$. 
 \\

  The above remains true in a full string theory embedding. Gravitons 
  are now the zero modes of the closed strings.  The fundamental parameters of the theory are the string scale, $M_s$, and a dimensionless string coupling, $g_s$.  In our discussion, we shall always stay within the domain of a weak string coupling, $g_s \ll 1$.\\  
  
  Let us assume that in string theory we somehow managed to construct a de Sitter space 
 with  a curvature radius $R_{dS}$.
 Intuitively, it is clear that if there exists any sensible notion of 
  de Sitter {\it geometry}, it must have a curvature radius $R_{dS}$ larger than 
  the string length $1/M_s$. This intuition is supported by the fact that 
 for $R_{dS} \ll 1/M_s$ the Gibbons-Hawking 
 temperature \cite{Gibbons:1977mu} of the de Sitter space 
 would exceed the string scale.  At such high temperatures  
  the Hagedorn effects \cite{Hagedorn:1965st} must become 
 important due to an exponentially growing number of active 
 degrees of freedom. Therefore,  
 irrespective whether the temperatures above $M_s$  are reachable in 
 string theory (see, e.g., \cite{Bowick:1992qu}),  it is 
 beyond any reasonable doubt that for such a state 
 the standard geometric description
 of classical GR must be abolished. 
  Therefore, without loss of generality, we can assume 
  that if de Sitter exists in string theory, its curvature must be below 
  the string scale.  \\

  Next, we wish to promote the assumed de Sitter
  into a quantum vacuum.  For this, we must take the 
  double-scaling limit (\ref{GLimit}).  
  For example, 
  to get a would-be
   de Sitter vacuum in ten dimensions, 
  we must take the ten dimensional Newton's constant
  $G_{(10)}$ to zero  while keeping $R_{dS}^{-2} = \Lambda G_{(10)}$ finite. 
  But, in such a limit, the string coupling, 
  $g_s^2 = M_s^8G_{(10)}$, vanishes for any finite value of the string scale $M_s$.  Thus, just like in pure gravity, 
  in full string theory, even a hypothetical existence of de Sitter 
 vacuum is incompatible with the closed string scattering.  \\
 
 The origin of $\Lambda$ in the above example can be made slightly more explicit by invoking a construction in the spirit of \cite{Dvali:1998pa} which uses  (anti)$D$-brane tensions as the source of 
 de Sitter energy. 
   For example, let us, in  type $II$B theory, pile up $n$ pairs of $D_9$-branes and their anti-branes,  $\bar{D}_9$, 
 on top of one another.  The resulting cosmological term 
 $\Lambda$ 
 is given by the sum of the brane tensions,
 \begin{equation} \label{radius} 
    R^{-2}_{dS} = n g_s M_s^2.
    \end{equation}  
  As a consistency check, notice that the requirement that 
 de Sitter temperature is less that $M_s$, 
 puts a restriction $g_sn < 1$.  This is identical to the constraint  
 on the number of Chan-Paton species 
 \cite{Dvali:2010vm}, 
 \begin{equation} \label{CP} 
 N_{\rm species} = n^2 < 1/g_s^2 \,.  
     \end{equation}   
 This expression represents as a special case of the
 general black hole bound on the number of particle species 
in a $d$ dimensional theory with a gravitational cutoff scale $M_{*}$ 
  and Newton's constant $G_{(d)}$ 
    \cite{Dvali:2007hz}, 
   \begin{equation} \label{BoundN} 
  N_{\rm species} <  \frac{1}{M_*^{d-2}G_{(d)}} \,.
    \end{equation}

    For $M_{*} = M_s$ in $d=10$ this gives, 
   \begin{equation} \label{BoundN10} 
  N_{\rm species} <  \frac{1}{M_s^{8}G_{(10)}} = \frac{1}{g_s^2} \,.
    \end{equation}  
    \\  
    
 The non-gravitational world-volume physics of $D_9 - \bar{D}_9$ is 
 sufficiently well understood \cite{Srednicki:1998mq} and the system is known to be unstable  because of the open string tachyon. 
 This instability leads to brane-anti-brane annihilation which can  
be described as tachyon condensation, studied in 
series of papers (see, e.g., \cite{Sen:1999mg}). \\
 
 Nevertheless,  the above brane-anti-brane configuration can 
 produce a classical ``hill-top" de Sitter in the limit $g_s \rightarrow 0$, while keeping $R_{dS}$ finite. 
  Again, we see that in the limit in which we can talk about a 
  classically-eternal de Sitter ``vacuum", the strings become non-interacting.  So are the zero mode gravitons. \\
 
 The above simple argument encodes a profound point. 
 It shows that a cosmological background can serve as an 
 eternal de Sitter vacuum but only at the expense of trivializing quantum gravity. 
This exposes a fundamental tension between de Sitter vacua and 
the $S$-matrix description of quantum gravity.  \\
  
 \subsection{De Sitter as Saturated Coherent State} 
 
  A physics common sense should be telling us that there must exists   
a microscopic reason behind this tension.
In order to identify it, we shall treat de Sitter as state of gravitons 
on Minkowski \cite{Dvali:2011aa}.  
More precisely, we represent de Sitter as coherent state, $\ket{dS}$,  constructed on top of a Minkowski vacuum  \cite{Dvali:2013eja, Dvali:2014gua, Dvali:2017eba}. 
   The choice of a coherent state is dictated by its maximal classicality.  Other corpuscular resolutions of de Sitter give similar results \cite{Dvali:2013eja} (see, \cite{Kuhnel:2014gja} for various implementations 
of this proposal).  \\
   
  Now,  since Minkowski is a valid $S$-matrix vacuum, this 
  approach seemingly avoids the problem with constant 
  $\Lambda$. For example, the same  $2 \rightarrow 2$ graviton scattering 
  process now can be viewed as an $S$-matrix process, 
 \begin{equation} \label{22}
  {\rm dS + grav_1 + grav_2 \rightarrow dS' + grav_1' + grav_2'} \,, 
  \end{equation}
  during which the de Sitter coherent state changes, 
 $\ket{\rm dS} \rightarrow \ket{\rm dS}'$.  \\

This however does not save 
 de Sitter because, even in the absence of any external gravitons, 
  the coherent state
 $\ket{\rm dS}$ time-evolves on its own.  This evolution 
leads to a breakdown of classicality after a certain  time $t_Q$. 
This phenomenon is referred to as {\it quantum breaking}, 
  with  the corresponding {\it quantum break-time} $t_Q$. 
  The time $t_Q$ marks a point of a complete departure
 of the quantum evolution from the classical one.          
 The concept was introduced in \cite{Dvali:2013vxa}  
   and was generalized to de Sitter, inflation and black holes  
 in  \cite{Dvali:2013eja, Dvali:2014gua, Dvali:2017eba}.
The outcome of this analysis is that a classical source of a de Sitter type metric must evolve faster than the corresponding $t_Q$.   
 In the opposite case, the source cannot be 
 embedded in a quantum theory. 
 In particular, this constraint excludes a constant $\Lambda$. 
 We shall briefly recount the main ingredients of the story
 required for the present discussion.
 \\

 First, viewed as a coherent state, the de Sitter Hubble 
 patch represents a so-called {\it saturated} state.
 Such is a state in which the mean occupation number
 ($N$) of dominant constituents and their quantum coupling ($\alpha$)
 satisfy the following criticality relation \cite{Dvali:2011aa}, 
  \begin{equation} \label{Crit1}
 N = \frac{1}{\alpha} \,. 
  \end{equation} 
  In the coherent state picture of de Sitter, there 
exist two types of constituents. 
 The first category originates from  the composition of the 
 gravitational field. 
  This category is universal and is present already  
  in pure gravity sourced by a constant $\Lambda$. 
  The second category is represented by non-gravitational quantum constituents of the source, provided the source is dynamical.  
 For example, the  role of such a source can be played by an inflaton potential or by a $D$-brane tension, as in \cite{Dvali:1998pa}. \\
  
 Regardless of the nature of the source, the gravitational constituents have the characteristic 
 frequency $\sim R_{dS}^{-1}$. Their quantum gravitational coupling 
 and the occupation number are related through the saturation relation, 
 \begin{equation} \label{Crit}
 N= \frac{1}{\alpha_{gr}} = \frac{R_{dS}^{2}}{G} \,. 
  \end{equation}

  In general, it is important to understand that the coherent state resolution of de Sitter (or of any other entity) is largely independent of 
 UV-completion.  Rather, it 
  amounts to a {\it corpuscular} resolution of the state.  Such a description probes the IR scales corresponding to 
  the size of the object.   
 This size, in general, is much larger than the
  cutoff scale. Therefore,  if a corpuscular resolution
  exhibits some inconsistency,  this in general cannot be cured by UV-physics.  Because of this, a coherent state view  of de Sitter ``vacuum"  is a convenient tool for monitoring  
  its compatibility with the $S$-matrix formulation. \\ 
  
   Notice, since, as discussed above,  the de Sitter curvature is less than the string scale, 
 the gravitational constituents of the metric are represented by the 
 zero mode sector of closed strings.  
 That is, 
 the resolution of the metric is mainly controlled by 
 IR gravitons. The contribution from the
 closed string states of mass $m$ 
 is exponentially small, $\sim {\rm exp}(-mR_{dS})$. 
  \\

  Next, we observe that the corpuscular picture gives a
 well-defined microscopic meaning to the rigid limit  (\ref{GLimit}). In the language 
  of a coherent state, this limit translates as, 
   \begin{equation} \label{NLimit}
 N = \frac{1}{\alpha_{gr}} \rightarrow \infty, 
 ~~~ R_{dS} = {\rm finite}\,. 
  \end{equation} 
   That is, de Sitter remains a saturated state but the number of 
   its constituents becomes infinite. 
   The corpuscular picture allows for a microscopic 
 visualization of how de Sitter is deformed into a valid vacuum 
 in the limit (\ref{GLimit}) and (\ref{NLimit}).
   \\

   In order to see this, let us consider the previous 
   $2 \rightarrow 2$ scattering process 
   of gravitons on de Sitter background (\ref{22}).  
 The meaning of the change 
 $\ket{\rm dS} \rightarrow \ket{\rm dS}'$
 is that the inner {\it corpuscular} structure of de Sitter is affected. 
 This is 
 due to interactions between the scattered gravitons  and the constituents of the de Sitter coherent state.  The effect 
 of these interactions can be split in two parts:  \\
 
 1) {\it A collective recoil.}   During this process, the recoil momentum is delivered to the entire coherent state,  without exciting its individual 
 graviton constituents.  This phenomenon can be viewed 
  as something like a {\it gravitational M\"ossbauer effect}.  \\
  
  2) {\it A corpuscular recoil}. During this process, the recoil 
 excites some individual quanta of the coherent state. This impact
 is suppressed 
  by the gravitational coupling (\ref{Crit}), and 
  therefore amounts to $1/N$ effect.   \\
  
   Due to the above two effects,  the de Sitter coherent state
 is altered by the scattering process.  
   However,  both 
   back reaction effects 
   vanish in the limit (\ref{NLimit}). First, the collective recoil is uniformly shared by
   an infinite number of quanta.  Such a state recoils, without 
   being affected. 
    Secondly, the recoil by the individual 
   constituents, since it is suppressed by $1/N$, also vanishes. 
    This makes it clear why de Sitter becomes a rigid vacuum 
   exclusively in the double-scaling limit (\ref{GLimit}), or equivalently, 
   for (\ref{NLimit}).  \\
   
   This picture also makes transparent the special role of Minkowski 
   as of $S$-matrix vacuum. Indeed, the Minkowski space corresponds 
   to de Sitter with an infinite curvature radius $R_{dS} = \infty$. 
   Correspondingly, viewed as the coherent state of gravitons, Minkowski  is composed out of 
   $N = \infty$ gravitons \cite{Dvali:2015rea}.  These gravitons have  zero frequencies
   and infinite wavelengths.  Hence, their gravitational couplings  
 $\alpha_{gr}$ vanish even for finite $G$.   
   Because of this,  
    both recoils vanish. 
    This is why, unlike de Sitter, the Minkowski space is 
   a valid $S$-matrix vacuum regardless of the value of $G$.  \\ 
   
   \subsection{$1/N$ Effects}
   
   We are now approaching a crucial point. 
   Namely, the coherent state resolution of de Sitter likely
   exposes its intrinsic inconsistency.  
   Due to a perpetual {\it inner re-scattering} among its constituents, the de Sitter coherent state looses its classical properties. 
  Note, for other systems a departure from classicality is not necessarily 
  an inconsistency. The conflict for de Sitter is that it is sourced 
  by a constant which remains permanently classical. \\ 
  
   The potentially-deadly effects are only of order $1/N$ per Hubble time,   
    but they are persistent and give a dramatic effect over a longer period.  
   It takes time $t_Q \sim NR_{dS}$ (\ref{DSS}) for the tension 
   to build up.    
     After this time,  the ``vacuum" evolves into a state 
   that is incompatible 
   with the constant classical source $\Lambda$ that produced 
   de Sitter in the first place.  This is an 
   inevitable conflict because the $1/N$ interactions cannot be 
   switched-off.     \\
   
 As a by-product of the inner re-scattering, the de Sitter coherent state 
  permanently emits particles and depletes. The emitted particles are nothing but the 
   celebrated Gibbons-Hawking radiation \cite{Gibbons:1977mu}. 
   The corpuscular picture is telling us that the radiation comes from an actual decay 
   of the coherent state \cite{Dvali:2013eja}.  \\
   
   Notice, Gibbons-Hawking radiation is controlled by the collective coupling 
   $\alpha_{gr}N \sim 1$.  It therefore survives in the rigid 
 limit  (\ref{NLimit}). In this limit, Gibbons-Hawking spectrum becomes 
 exactly thermal. At the same time de Sitter becomes an eternal vacuum. The problem, as already explained, is that this limit 
 cannot describe an interacting quantum gravity or strings. 
  A non-trivial quantum gravity requires a finite $N$. 
   Then, the  anomalous $1/N$ corrections  
 give deviations from the thermal spectrum. At the same time, they violate classical de Sitter invariance of the background. 
 These effects vanish 
 for (\ref{NLimit}) but are present for any finite $N$. 
 In other words, we can say that de Sitter is {\it anomalous}
with respect to $1/N$ quantum effects. 
 \\
 
In order to avoid misuse of the language, let us explain 
better the meaning of the above statement. Usually, anomaly 
implies a breaking of a classical symmetry by quantum corrections.
Once de Sitter is described as a coherent state on Minkowski vacuum, 
the symmetries of the vacuum are symmetries of Minkowski. 
The theory of course has a gauge redundancy in form of the general 
covariance.  What is affected by $1/N$ effects is the symmetry of a coherent state $\ket{dS}= \ket{N}$. For an observer that is blind to $1/N$ corrections, 
this coherent state is mistaken for a vacuum. This  ``vacuum" 
is invariant under a  symmetry that the observer calls a de Sitter group.  However, this classical picture is only an approximation 
that emerges in the 
limit (\ref{NLimit}). \\ 

  In order to visualise the dynamics, let us oversimplify the story by focusing 
 on a particular decay process. Let the theory contain 
 a set of light particle species, with the masses $\ll 1/R_{dS}$.    
  Then, an off-shel constituent graviton of energy $\sim 1/R_{dS}$ 
  can decay into a pair of such particles. 
  Of course, the decay is democratic due to the universality of graviton 
  coupling.  \\
  
  The rate of the process scales as
  $\Gamma \sim  (\alpha_{gr} N)/R_{dS}$ times the number of species. 
  So, the 
  coherent state looses roughly one constituent  per time 
  $R_{dS}$.  The produced quanta are a part of Gibbons-Hawking 
  radiation which is democratic in species.  \\  
  
   As a result of such a decay, the de Sitter coherent state 
 will time-evolve 
 into a (in general entangled) superposition of the sort
 \begin{equation} \label{j} 
  \ket{N} \rightarrow \sum_j c_j \ket{N-1}_j\times \ket{j}  
\end{equation} 
 where $c_j$ are coefficients and index $j$ runs over various states of the 
 produced quanta and the back-reacted de Sitter.     
  Of course, there exist a lot more processes contributing to the evolution 
  and generation of the entanglement but this one suffices to explain the point (see, \cite{Dvali:2013eja, Dvali:2017eba} for more technicalities).     
  \\
  
  Now, an observer calls the initial state  $\ket{N}$ a de Sitter because 
  the expectation value of a graviton field over it gives a classical de Sitter 
  metric. After the decay, this expectation value is going to change 
  in the way that departs from the classical evolution and 
  thus from its classical symmetry. In this sense, a more precise 
  term would be to say that the classical de Sitter symmetry
  is {\it spoiled} rather than broken. The difference is that in the second case, 
  one should be able to define a time-evolution which, while 
   different from de Sitter, is still classical.  Nothing like this is 
   feasible because the classicality of the state is affected. 
   That is, $1/N$ effects break the classical symmetry of the background because they 
   abolish its classicality. \\
   
    Notice, the above is true, even if we imagine that
   a symmetry generator, call it $\hat{T}$, that annihilates the state 
   $\ket{N}$, commutes with the time-evolution operator that is responsible  
   for the transition (\ref{j}). Of course, in such a case  
   $\hat{T}$ also annihilates the r.h.s. of (\ref{j}).  So 
   the evolved state has the same symmetry as the original one. 
   However, This does not imply that the background remains unchanged. 
   In fact, it is unclear which state should be called a new de Sitter 
   vacuum. It is tempting to say that  this is $\ket{N-1}_j$, which is obviously 
   different from $\ket{N}$. However, even this  choice is not clean, since 
   $\ket{N-1}_j$ carries a quantum index through which it is entangled with the rest.

 \subsection{Quantum Break-Time}

In anomalous quantum breaking of de Sitter, its saturation 
close to criticality (\ref{Crit})   
 is an important factor.  
  While some departure from the classicality 
  is experienced by other macroscopic systems, the saturated ones 
  are special, as they can depart {\it fully}.   
Without repeating technicalities that can be found in 
\cite{Dvali:2013eja, Dvali:2014gua, Dvali:2017eba,Dvali:2013vxa},
below we display some useful equations.
 \\

  For a generic saturated system, in the absence of a classical Lyapunov instability,  the quantum break-time is given by
  the following formula \cite{Dvali:2017eba},    
  \begin{equation} \label{QBT}
    t_Q = \frac{t_{\rm cl}}{\alpha N_{\rm species}} = t_{\rm cl} 
    \frac{N}{N_{\rm species}} \,,
    \end{equation}  
where $\alpha$ is the quantum coupling and $t_{\rm cl}$ is 
a characteristic time on which classical non-linearities 
become important.  As already introduced previously, $N_{\rm species}$ counts the number of 
the active light particle species in the spectrum of the theory, that interact  
via coupling $\alpha$. 
 This number should not be confused with the occupation number 
of quanta $N$.
\\

Notice that, generically,  the point $\alpha N_{\rm species} =1$ 
marks a non-perturbative saturation of unitarity by the species \cite{Dvali:2020wqi}.
In this respect, the bound (\ref{BoundN}) represents a particular manifestation of this phenomenon in case of gravity.  
Thus, the equation (\ref{QBT}) tells us 
that, in an unitary system, the quantum break-time 
$t_Q$ cannot be shorter than the classical time
$t_{\rm cl}$. \\

Now, the systems that are subject to classical instabilities, 
are special from quantum breaking point of view.  
If a system exhibits a classical Lyapunov time $t_{\rm cl} =t_{\rm Lyapunov}$, the quantum break-time can be as short as \cite{Dvali:2013vxa}, 
  \begin{equation} \label{QBTL}
    t_Q = t_{\rm Lyapunov}\ln(\alpha^{-1}) = t_{\rm Lyapunov}\ln(N) \,.
    \end{equation}  \\

 Thus, a classical instability can speed up the process of 
 quantum breaking.   
However, in a generic system, the quantum break-time  
is not necessarily connected with instability 
but always signals a breakdown of the classical 
description.   
 \\

\subsection{$S$-Matrix and Quantum Breaking} 
 
 Using a proper expression for $t_Q$, it is easy to follow how the $S$-matrix constraint goes hand in hand 
 with the quantum break-time criterion. 
 For example, in $d=4$ theory of 
pure gravity, we must take  $t_{\rm cl} = R_{dS}$
 and  $\alpha = \alpha_{gr} = G/R_{dS}^2$. Then, from 
 (\ref{QBT}),  the  resulting quantum break-time  is 
\cite{Dvali:2013eja, Dvali:2014gua, Dvali:2017eba},   
\begin{equation}\label{QBTE}
  t_Q = \frac{R_{dS}^3}{G} \,.
 \end{equation}   
The de Sitter vacuum is only possible 
if this time is taken infinite.   This requires, $G \rightarrow 0$. 
But then, the gravitational $S$-matrix becomes trivial.  \\

  The existence of additional particle species, changes the story. 
  According to (\ref{QBT}), the quantum break-time is shortened   
   if the theory contains a large number 
   $N_{\rm species}$ of light particle species.  In this case the 
   equation (\ref{QBTE}) must be replaced by, 
\begin{equation}\label{QBTSp}
  t_Q = \frac{R_{dS}^3}{GN_{\rm species}} \,.
 \end{equation}
Taking into account the black hole bound (\ref{BoundN}) on 
$N_{\rm species}$, we can convert the above equation 
into the following bound, 
 \begin{equation}\label{QBTM}
  t_Q >  R_{dS}(R_{dS}M_*)^2 \,.
 \end{equation}
 This expression tells us that, as long as $R_{dS}$ is larger than the cutoff length 
 $1/M_*$, the quantum break-time of de Sitter is 
 longer than the Hubble time.  \\
    
 The existence of particle species has important effect on gravitational physics,
 both via black holes and via cosmology. 
 This is already clear from the fact \cite{Dvali:2007hz} that species lower the gravitational 
 cutoff of the theory $M_*$ (relative to Planck mass 
 $\equiv 1/G_{(d)}^{{1/(d-2)}}$)  
  according to (\ref{BoundN}).  
 The equations (\ref{QBTSp}) and (\ref{QBTM})  
 reveal that $N_{\rm species}$-dependences 
of $M_*$ and of $t_Q$ are the two sides of the same coin. 
This is because no sensible semi-classical state can 
quantum break during the time $t_Q$ shorter than (or comparable to) its classical time $t_{\rm cl}$.  As it is clear from (\ref{QBT}), the increase of the number of species, speeds up the quantum breaking process and moves $t_Q$ closer to $t_{\rm cl}$.  A system for which
$t_Q \sim t_{\rm cl}$, cannot be described within an unitary effective theory.
For such  a system, $t_Q$ marks the cutoff length.   
\\

 In order to highlight the important role of species, let us
  take another 
 double-scaling limit, 
   \begin{equation}\label{LSpecies}
   G \rightarrow 0, ~ N_{\rm species} \rightarrow \infty,~
   GN_{\rm species} = {\rm finite}, ~R_{dS} = {\rm finite} \,.
 \end{equation}
 In order to keep $R_{dS}$ finite, this must be combined with 
 the previous large-$N$ limit (\ref{NLimit}) (or equivalently, with (\ref{GLimit})).   
 The limit (\ref{LSpecies}) is also analogous to 't Hooft's planar limit in QCD. 
 The role of the number of colors in gravity is assumed by the 
 total number of particle species $N_{\rm species}$.  \\
 
 One interesting thing about the above limit is that, despite the fact that the gravitational coupling vanishes, the quantum break-time 
 (\ref{QBTSp})
 remains finite. 
This is due to the collective effect of species.
Thus, in the combined limit (\ref{LSpecies}), de Sitter never becomes a rigid vacuum. \\

This phenomenon can be understood in two languages. First, 
since $R_{dS}$ is finite, one naively expects to have a constant
Gibbons-Hawking temperature $T_{GH} = 1/R_{dS}$.  However, this is not possible since, under the assumption of an exact thermality,  
the intensity of Gibbons-Hawking radiation  would diverge 
due to infinite number of light species. Thus, 
the back reaction remains non-zero and must be taken into account. \\

The second way to understand the same effect is from 
the point of view of the coherent state picture of de Sitter
\cite{Dvali:2013eja, Dvali:2014gua, Dvali:2017eba}. 
The radiation is a result of the depletion of the coherent state. This is
due to decays and various re-scatterings of its constituents. 
For example, the simplest processes are the  
direct decays of constituent off-shell gravitons, or their pair-wise annihilations
into species.  As already noted, due to universality of 
the gravitational interaction, the particle creation is democratic in species.  
This explains the universality of Gibbons-Hawking radiation. 
The resulting half-decay time of the $N$-graviton coherent state is  
\begin{equation}\label{QBTSpN}
  t_{decay}  = R_{dS} \frac{N}{N_{\rm species}} \,.
 \end{equation}
Taking into account (\ref{Crit}), this expression exactly matches 
(\ref{QBTSp}). \\

 So far, we did not specify the internal dynamics of the 
 energy density source that produces the de Sitter state. 
 Of course, for constant $\Lambda$, there is no dynamics and 
 the entire quantum breaking process is due to gravity.  
  The finiteness of $t_Q$ reveals the inconsistency of $\Lambda$. 
  Thus, a valid source must change in time, in order to avoid 
  a conflict with quantum break-time.  Such a source, is expected 
  to have its internal non-gravitational dynamics. 
  An example is a scalar field with a potential that 
  asymptotes to zero.  \\ 
   
 Here we must distinguish among the two types of processes that contribute 
 to quantum breaking: 
 1) Re-scattering of the constituent gravitons;   
 and 2) re-scattering of the constituents of the source. 
  The first mechanism is generic and is independent of the composition of the source.  It therefore provides an universal constraint. \\
  
  However, the quantum breaking effect can independently 
be exhibited by a non-gravitational dynamics of the source.  
The corresponding quantum break-time is expected to obey 
(\ref{QBT}), where the parameters $\alpha, N, t_{\rm cl}$
must be understood as the non-gravitational characteristics 
of the source.  For example, $N$ is the occupation number of 
a scalar field in the coherent state, $\alpha$ is its self-coupling  
and $t_{\rm cl}$ is the corresponding classical time 
(see, \cite{Dvali:2013eja, Dvali:2017eba, Dvali:2017ruz}
for applications of quantum breaking 
to some inflationary scenarios and 
other systems.)   
 \\
   
\subsection{$t_Q$ in String Theory} 
Let us now move to string theory. 
 Let us assume that a would-be de Sitter  
 of curvature radius $R_{dS}$  is achieved by some 
 construction.  Since $R_{dS}$ sets the wavelengths/frequencies  
 of graviton constituents of the de Sitter coherent state, 
  their quantum coupling is given by, 
   \begin{equation} \label{Astr} 
   \alpha_{gr} =  \frac{G_{(10)}}{R^{8}_{dS}} =
   \frac{g_s^2}{(R_{dS}M_s)^8}\,.
    \end{equation}  
 Plugging this into (\ref{QBT}) and taking into account 
  $t_{\rm cl} = R_{dS}$, we obtain the following expression 
  for quantum break-time, 
     \begin{equation} \label{Qstr} 
   t_Q = R_{dS}  \frac{(R_{dS}M_s)^8}{g_s^2} \frac{1}{N_{\rm species}}\,.
    \end{equation}  
   Here, $N_{\rm species}$ is the number of light particle species 
   with masses below $1/R_{dS}$. The example shall be given below. 
 It is obvious that for finite values of 
 $R_{dS}$ and $M_s$, the only way of making $t_Q$ infinite 
 is by taking $g_s =0$ while keeping  $g_s^2N_{\rm species}$ finite. 
   This renders the closed string 
 $S$-matrix trivial.  \\
 
 Since the expressions (\ref{Astr}) and (\ref{Qstr}) make no assumptions 
 about the nature of the source,  they set an universal upper bound on the 
 quantum break-time. This upper bound is sufficient for excluding 
 de Sitter vacua from string theory. However, in particular cases, 
 especially when the source exhibits a Lyapunov exponent, the quantum breaking of the source itself could happen much faster.
 Of course, such an instability only adds to the tension 
 against de Sitter.  
  However, the gravitational contribution (\ref{Qstr}) to $t_Q$, 
  persists regardless of other effects. 
    \\ 
   
 For example, 
 let us come back to the system with $n$ pairs of 
 $D_{(9)}-\bar{D}_{(9)}$-branes piled up on top of each other. 
 In this case, the curvature radius is given by (\ref{radius}). 
  From (\ref{Astr}), the corresponding quantum coupling of the graviton constituents of  
 de Sitter is, 
 \begin{equation} \label{AstrD} 
   \alpha_{gr} =  
    (n g_s)^4g_s^2\,.
    \end{equation}  
 Taking into account the number of Chan-Paton species,
 $N_{\rm species} = n^2$, and using (\ref{QBT}), we get the following quantum break-time, 
    \begin{equation} \label{QstrD} 
   t_Q = \frac{1}{M_s}\frac{1}{(n g_s)^{13/2}} = 
   \frac{R_{dS}}{(ng_s)^6}\,.
    \end{equation}
   As another consistency check, notice a  
   quartic agreement. 
  First, the requirement 
   that the de Sitter quantum break-time 
   is less than the Hubble time, $t_Q <  R_{dS}$, puts the bound 
   on the number of Chan-Paton  factors, given by (\ref{CP}). 
   This is exactly the same bound as was derived 
   in \cite{Dvali:2010vm} by imposing the black hole
   bound  \cite{Dvali:2007hz} (\ref{BoundN}) 
   on Chan-Paton species. 
   Next, as noticed earlier, 
   the same bound comes from the requirement that 
   de Sitter temperature is less than the string scale.  
    Finally, according to \cite{Dvali:2020wqi}, the same bound 
  is imposed by the unitarity of 
  Chan-Paton scattering amplitudes.  \\

  Now, the quantum break-time due to instability  of the source, 
  is much shorter.    
Indeed,
taking into account the Lyapunov exponent originating 
from the open string tachyon
$t_{\rm Lyapunov} \sim M_s^{-1}$ and  using (\ref{QBTL}), we get, 
 \begin{equation} \label{stringyL}
t_Q = \frac{1}{M_s}\ln{(g_s^{-2})} \,.   
\end{equation} 
This type of quantum breaking can in principle be avoided 
by assuming that the $D$-brane system somehow got stabilized
in a local minimum.   Of course, this would kill the Lyapunov exponent and would eliminate (\ref{stringyL}).
We are ready to grant such a possibility.  
 \\
   
  However, regardless of this assumption, there is no 
  visible way to eliminate (\ref{Qstr}), as it comes from the 
  gravitational re-scattering of finite anergy quanta.  
  Therefore,  this expression 
  sets an upper bound on quantum break-time of an arbitrary de Sitter like 
  state in string theory. \\ 
  
  \subsection{Connection to Witten-Veneziano}

 In the picture of \cite{Dvali:2013eja, Dvali:2014gua, Dvali:2017eba}, 
 which we follow, de Sitter is not a vacuum but rather a coherent 
 state built on top of Minkowski. So, the de Sitter symmetry is not a property of a vacuum but an effective (emergent) symmetry of a particular coherent 
 state with $N$ constituents. This symmetry becomes exact in the limit 
 of infinite $N$ (\ref{NLimit}) which is equivalent  to (\ref{GLimit}).  
 For a finite $N$, it is abolished by 
 $1/N$ quantum corrections. \\

 The anomalous decay of de Sitter due to $1/N$ effects  
 cries for establishing analogy 
  with the breaking of axial symmetry 
 by  Witten-Veneziano mechanism in QCD with $N$-colors
 \cite{Witten:1979vv}. 
   Without any reference to fundamental fermions, this theory can be
   viewed as theory of mesons and glueballs. 
   They interact 
  via a quantum coupling that has $1/N$ strength. 
  The theory also includes baryons with mass $\sim N$.
   Just like in the corpuscular picture of de Sitter, the  
  classical limit corresponds to $N=\infty$.  In this limit there 
  exists  an exact 
  shift symmetry of one of the mesons ($\eta'$), a Goldstone bosons
   corresponding to a non-linearly realized axial $U(1)$.
  However, at the quantum level this symmetry is explicitly 
  broken by $1/N$ effects which generate mass and a 
  finite lifetime for the $\eta'$-meson. 
  In the certain sense, this is similar to how the quantum $1/N$ effects 
  generate a finite quantum 
  break-time for de Sitter.  However, unlike the $\eta'$-meson, the 
  quantum breaking of de Sitter appears to be in conflict with its classical source,
 unless the source evolves in time. \\

\subsection{Some Comparisons} 

  It is useful to comment on how the conjectures by other authors 
 fit within the presented framework.  
  We shall start with the one 
 by Banks \cite{Banks:2000fe}. While this work contains several points 
on which we have no immediate baring, one sharp conjecture made there 
is that de Sitter Hilbert space has a finite dimensionality. 
This statement 
can be taken with some skepticism. 
 However, the present picture offers an interpretation
 and possibly some support.  \\
  
  As discussed,  our view is that in an $S$-matrix theory of gravity 
  de Sitter cannot be a vacuum. Instead, it must be viewed  
as a composite \cite{Dvali:2011aa} coherent state in the Hilbert space 
built on top of the $S$-matrix vacuum of Minkowski 
\cite{Dvali:2013eja, Dvali:2014gua, Dvali:2017eba}.  
 Of course, the entire Hilbert space is infinite with de Sitter being 
 a particular composite state in it. 
 Nevertheless,  to the notion of finiteness of the ``de Sitter's Hilbert
 space " can be given a well defined meaning. 
  We can attribute it to a dimensionality of the portion of the Hilbert space 
  explored by the state vector over the quantum break-time $t_Q$.
  This is a meaningful prescription, since beyond $t_Q$ the state vector does not describe anything close to a classical de Sitter vacuum.       
   \\ 
   
 A number of authors (see, e.g., Tsamis and Woodard \cite{Tsamis:1996qq},  
Polyakov  \cite{Polyakov:2012uc} and
 Anderson, Mottola and Sanders \cite{Anderson:2017hts}) have suggested instabilities of de Sitter 
 space.  These proposals are very different (both from us and also from each other) but touch a certain common point.
We share the point (especially with the approach by Polyakov)
that de Sitter cannot keep producing particles  ``for free"
and some price must be payed. 
The difference is that in these papers 
 it is envisaged that, due to back reaction, either 
 de Sitter is destabilized or the curvature of de Sitter space 
 is decreasing in time. 
 Basically, effectively $\Lambda$ is getting screened as the time goes on.
  This is not necessarily in contradiction to our findings but represents 
  a fundamentally different claim. \\
  
   The picture \cite{Dvali:2013eja, Dvali:2014gua, Dvali:2017eba} that we advocate is 
  not of a decrease of the classical curvature (or screening
  $\Lambda$) in time. It is in principle impossible to
 describe the quantum evolution of the de Sitter coherent state, 
 triggered by $1/N$ effects, in 
 terms of time-dependent classical characteristics such as the 
 curvature. Rather,  $1/N$ effects induce a departure from the classical 
 description and its complete 
 invalidation after the time $t_Q$.  
  Therefore, in our picture, the de Sitter with 
  a constant $\Lambda$ is {\it inconsistent} rather than {\it unstable}. \\
  
  Of course, if different quantum gravity effects happen 
  to screen $\Lambda$ 
  on times shorter than $t_Q$, this may help 
  in avoiding the inconsistency. {\it A priory} we have nothing  
  against such a possibility but it is beyond our claims. \\
  
  Finally,  some constraints on the scalar potentials 
 that speak against de Sitter, were also conjectured recently 
  \cite{Obied:2018sgi},\cite{Ooguri:2018wrx}.
  As discussed in \cite{Dvali:2018fqu}, these constraints 
 are  essentially equivalent to the ones obtained from the quantum break-time \cite{Dvali:2013eja, Dvali:2014gua, Dvali:2017eba}, for a 
 special choice of $\alpha \sim 1$.

\subsection{Observational Signatures and Power of Species} 

 We would like to briefly outline some observational 
signatures that follow from our approach. Since our work is
a continuation of the framework \cite{Dvali:2013eja, Dvali:2014gua, Dvali:2017eba},  part of the signatures 
 have already been discussed there.  For completeness, we 
 shall outline the main directions and then discuss some new 
 ideas. \\

   The first obvious prediction, relevant for the present day cosmology,
   is the exclusion of a constant $\Lambda$ from the energy density 
   budget in the Universe. 
    In this light, more precision tests of standard $\Lambda$CDM cosmology 
 \cite{Secrest:2020has} are called for. 
Our picture indicates that whatever contributes to the accelerated expansion of the Universe, cannot be $\Lambda$, or any other constant source.  We do not see, how such a source 
could be reconciled with the finite quantum break-time.
 \\
   
   Next,  a more indirect contact with observations 
  can be established  though 
  the quantum breaking constraints imposed on inflaton potentials. 
   The requirement that a scalar field must evolve faster than 
   the corresponding  $t_Q$ (\ref{QBT}),  puts a constraint on 
   the shape of its potential \cite{Dvali:2013eja}.   In a large class of theories, this effectively 
   translates as an upper bound on the inflationary slow-roll parameters 
   which limits the number of inflationary e-folds by, 
   \begin{equation} \label{NE}
      {\mathcal N}_e \lesssim \frac{1}{\alpha N_{\rm species}} \, .  
   \end{equation}  
   In particular, applied to string theory, we must take 
   $\alpha = g_s^2$.  \\

   Perhaps, the most interesting qualitatively-new observables 
are the corpuscular imprints from the 
   inflationary epoch \cite{Dvali:2013eja,Dvali:2018ytn}.
   The essence of the story is as follows. 
    The inflationary 
   paradigm is based on the assumption that the present Hubble patch 
   underwent through a  de Sitter epoch. In the semi-classical treatment, 
   there is no upper bound on the duration of this phase.  
  At the same time, the observable imprints are assumed to come only from last $60$ or so 
   e-folds. In the semi-classical picture, the entire information about the
  prior epoch is lost.  \\
  
  This is fundamentally changed in the corpuscular picture
  \cite{Dvali:2013eja}, 
 which is imposed upon us by the $S$-matrix consistency. 
This picture tells us that de Sitter possesses an intrinsic quantum clock. 
  This clock is powered by $1/N$ corpuscular effects. 
 After the time $t_Q$, this leads to a complete quantum breakdown of 
  the coherent state. Basically, one can say that de Sitter ``wears off". \\
    
   By consistency, we know that our 
  Hubble patch gracefully exited the de Sitter phase well before $t_Q$ elapsed.   That is, the exit time, $t_{exit} = {\mathcal N}_e R_{dS}$, 
  that measures the classical duration of inflation, must satisfy,   
  \begin{equation} \label{exit} 
    t_{\rm exit} \lesssim  t_Q\, .
 \end{equation}  
 
   Now, since the strength of the 
   corpuscular imprints is $\sim R_{dS}/t_Q$ per Hubble time, by the end 
   of inflation, the relative magnitude of the imprints is given by \cite{Dvali:2013eja},
  \begin{equation} \label{Aimprint1}
   \delta = \frac{t_{exit}}{t_Q} \,. 
\end{equation}      
  Thus, longer the inflation lasted, stronger are 
   the quantum imprints and higher are the chances of detecting them 
   via the precision cosmology.   \\ 

  A successful inflation must last longer than $60$ e-folds, in order to solve 
the horizon and flatness problems. 
 Taking this into account, we can write a combined bound, 
 \begin{equation} \label{ComB} 
   \frac{60 R_{dS}}{t_Q} <  \delta   < \epsilon_{\rm obs}\, , 
 \end{equation}  
where $\epsilon_{\rm obs}$  parameterizes the current observational 
accuracy. \\

  This brings us to the following point. 
 The existence of large number of particle species can 
 dramatically enhance the observable effect of quantum imprints. 
  This is revealed by the equation (\ref{QBTSp}), which is telling us that 
  species shorten $t_Q$.  Using the expression (\ref{QBTSp}),
  we can convert (\ref{ComB}) in an useful bound on the number of 
  inflationary  e-folds,   
   \begin{equation} \label{EIbound} 
  60 <  {\mathcal N}_e  <  \frac{N}{N_{\rm species}} \epsilon_{\rm obs} = 
 \frac{1}{N_{\rm species}}\frac{R_{dS}^2}{G}  \epsilon_{\rm obs} 
   \,,
 \end{equation} 
 where in the last equality we expressed $N$ through (\ref{Crit}). 
 At the same time, the relative strength  
 of imprints (\ref{Aimprint1}) takes the form, 
 \begin{equation} \label{imprints} 
 \delta \sim  {\mathcal N}_e  \frac{N_{\rm species}}{N} \,.  
\end{equation} 
 We see that the number of species 
 enhances the amplitude and simultaneously
 narrows the  window (\ref{EIbound}). \\
 
 In general, a scenario with $\frac{N_{\rm species}}{N} \sim 
 \frac{\epsilon_{\rm obs}}{60} $, 
 puts the effect near the observational accuracy. 
   This fact opens a large range of inflationary scenarios, 
   with different values of Hubble, for which the effect 
   can be observable. 
 Notice that already
 taking into account the 
existing number of particle species in the Standard Model, 
increases the effect by a factor of a hundred, as compared to 
the case with pure inflaton.    
 \\

 Of course, the best motivated are the theories in which the existence 
 of the large number of species is independently justified. 
 Many extension of the Standard Model (e.g., 
 grand unification), have this property.  We wish to 
 discuss an extreme case. This is a scenario
 with $N_{\rm species} \sim 10^{32}$  particle species, which is  motivated 
 by the Hierarchy Problem \cite{Dvali:2007hz}.  This number represents an 
 absolute phenomenological upper bound, since
 through (\ref{BoundN}) it brings the cutoff down 
 to $M_* \sim$ TeV.  Correspondingly, the Hierarchy Problem 
 gets nullified. 
 In a theory with such a large number of particle species, 
 the amplitude of corpuscular imprints can easily reach the 
 observable accuracy for the values of Hubble parameter 
 as low as $R_{dS}^{-1} \sim 10$GeV. \\

  In general, theories with many species open up a qualitatively new 
  way of generating observable quantum effects for 
  the values of the Hubble parameter for which the standard 
 inflationary fluctuations would be undetectable.  
 The species enhance the effect on two fronts. \\

 First, 
 their production due to decay of the coherent state, 
  increases the relative energy density of Gibbons-Hawking 
 radiation as compared to the energy of the inflaton background,
 \begin{equation} \label{Density} 
 \delta_{sp} =  \frac{N_{\rm species}}{N} = N_{\rm species}\frac{G}{R_{dS}^2} \,. 
  \end{equation}
   Secondly, due to back reaction on the de Sitter coherent state, the 
  species imprint the corpuscular corrections with the strength given by
  (\ref{imprints}).   \\
  
  In todays observations, the imprint (\ref{Density}) can be detectable 
   only from the species that were emitted from the coherent state during 
  the last $60$ e-folds. In contrast, the imprint of back reaction 
   (\ref{Density}) is cumulative due to 
  ongoing depletion since the onset 
   of inflation. This imprint carries a quantum information 
   about the {\it entire}  duration of inflation. It is remarkable how the  $S$-matrix consistency  correlates the two contributions.  \\   
    
 It is interesting that, for large values of  $N$ and $N_{\rm species}$,
 the contributions from species can  dominate over the standard contribution 
 from the fluctuations of the inflaton field.  
 The reason is that, although at large $N$ (small Hubble) 
all contributions are suppressed,  the imprints from the corpuscular effects, 
  (\ref{imprints}) and  (\ref{Density}), are enhanced by 
  the number of species $N_{\rm species}$, whereas the standard 
  inflaton  fluctuations are not. This can also be understood from the fact that
  in the limit 
  \begin{equation} \label{NewL}
  N \rightarrow \infty, ~~  \frac{N_{\rm species}}{N}={\rm finite} \,,
  \end{equation}
  the only surviving coupling is the collective coupling of species.
  All other interactions decouple.  
  Notice, that the limit (\ref{NewL}) is equivalent to 
  (\ref{LSpecies}) in which case gravity decouples while 
  $t_Q$ stays finite. The imprints from species encode  the information 
  about the finiteness of $t_Q$. 
   In such a regime, the fluctuations 
  enhanced by species can be a dominant source of primordial 
  quantum imprints in the Universe.  
   We thus observe that the theories with species can offer a possibility of probing the quantum sub-structure of the inflationary density 
 perturbations in a qualitatively different way.

\subsection{Outlook} 
 
  It is evident \cite{Dvali:2013eja, Dvali:2014gua, Dvali:2017eba} that 
  de Sitter vacua are inconsistent in quantum theory. 
  In this paper we have reconciled this view with the 
 $S$-matrix perspective.   
  We have observed how the quantum breaking of de Sitter vacua 
 is linked with their conflict with $S$-matrix. 
 This conflict is due to $1/N$ effects that 
 generate a departure from classicality.  A full quantum break takes place 
 after a finite time $t_Q$.   These effects vanish
 only in the limit of infinite $N$. However,  this limit implies $G=0$ and, in string theory,  $g_s=0$.  
  \\
 
 We have payed a special attention to the applications 
and cross-checks in string theory,   
as a prominent example of $S$-matrix theory. 
However, the presented  arguments are  very general and  should 
be applicable to an arbitrary $S$-matrix formulation of 
Einstein gravity, regardless of the UV-completion.
In particular, they are applicable to UV-completion by classicalization
\cite{Dvali:2010bf}.  
\\

   We converge to the following statement.  
  For fixed $R_{dS}$ and $M_s$, the quantum breaking 
   time of stringy de Sitter scales with $g_s$ as (\ref{DSS}). 
   It is unclear how to reconcile such a state with a classical 
   source that last longer than $t_Q$.  In such a case, 
a de Sitter like state in string theory,  irrespective 
of a seeming classical stability of the source,  can be neither stable
nor meta-stable.  
 The only possibility for obtaining an eternal de Sitter, is to make the quantum break-time infinite (while keeping $R_{dS}$ finite). This demands $g_s\rightarrow 0$, in which case the string $S$-matrix becomes trivial.  \\

 In general, a remedy for a de Sitter state is to asymptote to Minkowski vacuum sufficiently fast. 
 The term {\it sufficiently fast}  
 means faster than its local {\it quantum break-time},  $t_Q$. 
 In other words,  for $S$-matrix consistency, 
  the graceful exit time from de Sitter, $t_{\rm exit}$,
 must be shorter than the quantum break-time, (\ref{exit}). 
  \\  
 
 Naturally, the above puts severe constraints on de Sitter 
 model building.  
  Based on the idea that $D$-branes can be used for creating 
  de Sitter \cite{Dvali:1998pa}, a considerable effort went  in engineering such states in string  theory.  The  presented  $S$-matrix argument tells us that any such configuration 
  must evolve on the time-scales shorter than the corresponding 
  $t_Q$. \\

   As already discussed 
  \cite{Dvali:2013eja, Dvali:2014gua, Dvali:2017eba}, 
  the quantum break-time constraint has 
  number of consequences, 
   including the observational ones.
   First, of course, it
excludes a constant $\Lambda$ as part of the 
energy density in the Universe.  Whatever source contributes into 
the dark energy,  must be time-dependent.  \\

 Secondly, the presented picture point to new types of imprints 
 from the inflationary epoch \cite{Dvali:2013eja,Dvali:2018ytn}.
  These are the
 imprints from $1/N$ corpuscular effects that power the 
 quantum break-time clock.  The 
 strength of these imprints 
 is set by a parameter  $\delta = t_{exit}/t_Q$, also given 
 by (\ref{imprints}).  
 Correspondingly, the inflationary scenarios with short $t_Q$ are the most 
 interesting ones.  \\

 The number of 
  particle species, $N_{\rm species}$, plays a special role in shortening $t_Q$ 
and generating
  observable imprints from the inflationary epoch.
  Remarkably, through the quantum breaking constraint, the imprints 
  that come from last $60$ e-folds (\ref{Density}), are linked with 
  the ones that originate from the onset of inflation (\ref{imprints}).
 The latter imprints carry memories of the entire duration of inflation.  
     \\
    
   An important point is that the strengths of the corpuscular 
  imprints, (\ref{imprints}) and (\ref{Density}),  
  are enhanced by $N_{\rm species}$, while the standard infaton fluctuations, are not.
  Due to this,  species can provide a way of 
  generating the observable imprints even for the values of the Hubble parameter 
  for which the standard inflaton fluctuations are negligible.   
    Therefore, in  a wide range of the inflationary models, 
  the corpuscular effects, enhanced by the number of species, 
  can provide a major observational window in 
  the quantum substructure of the inflationary de Sitter. \\

 {\bf Acknowledgements} \\
 
 We thank Goran Senjanovi\'c for valuable discussions and comments.
Conversations on de Sitter $N$-portrait with Lasha Berezhiani, Cesar Gomez and Sebastian Zell are 
also acknowledged. 
This work was supported in part by the Humboldt Foundation under Humboldt Professorship Award, by the Deutsche Forschungsgemeinschaft (DFG, German Research Foundation) under Germany's Excellence Strategy - EXC-2111 - 390814868,
and Germany's Excellence Strategy  under Excellence Cluster Origins.

\end{document}